\documentclass[seceq]{ptptex}

\usepackage{wrapft}
\usepackage{hyperref}
\hypersetup{
    colorlinks=true,
    linkcolor=blue,
    filecolor=magenta,      
    urlcolor=blue,
    pdftitle={Weyl entropy behaviour and Naked Singularities},
    pdfpagemode=FullScreen,
    }

\title{
Weyl entropy behaviour and Naked Singularities\footnote{Based on presentation to be given at the 32nd meeting of the Indian Association for General Relativity and Gravitation, IISER Kolkata.}}    

\author{
\textsc{Vaibhav Kalvakota}$^{1}$\footnote{E-mail: vaibhavkalvakota@icloud.com}
}

\inst{
$^1$Turito Institute, Hyderabad, 500081, India\\
}

\recdate{
November, 2022}

\abst{
In order to show that there exists a relation between the validity of the Weyl entropy and the cosmic censorship hypothesis, we discuss the relation between naked singularities and the gravitational entropy estimator using the Weyl invariant. We look at the examples of Vaidya and Tolman-Bondi spacetimes in asymptotically flat and de Sitter spacetime and look at Weyl entropy as a possible indicator of the prohibition of naked singularities.
\vspace{3mm}
}

\begin{document}
\maketitle

\section{Introduction}
Following Penrose's Weyl curvature hypothesis \cite{a}, there has been a growing mathematical interest on the definition of gravitational entropy estimators and their cosmological implications (for instance Gron and Hervick\cite{b}, Chakraborty \textit{et al}\cite{c}, Pizana \textit{et al}\cite{d} and Lima \textit{et al}\cite{e}). The foundation of the Weyl hypothesis is that the Weyl tensor might be an estimator of the gravitational entropy of the spacetime. In using the Weyl tensor to define the gravitational entropy, there are three conditions to be met: \textbf{(1)} the Weyl tensor must be zero at $t=0$ for a cosmology and increase monotonically in the entire future evolution of the cosmology, \textbf{(2)} it must reduce to the holographic principle form for a given spacetime when applicable and \textbf{(3)} it must account for the anisotropies forming in the cosmology. Each of these points can be elaborated as follows: (1) implies that the gravitational entropy is monotonically increasing, accounting for structure formations in the cosmology. It must be zero at the initial singularity accounting for the perfect Friedmann-Robertson modelling of the spacetime, which has a metric that is conformally flat. Due to this, the Weyl tensor vanishes, which can be attributed to the lower anisotropies in the spacetime. Note that this does not correspond to the matter entropy, which clearly is non-zero. (2) refers to the nature of the gravitational entropy of spacetimes with a horizon, in which cases the gravitational entropy must reduce to the holographic principle. For black hole spacetimes this would be the Hawking-Bekenstein entropy, while for cosmologies with a horizon this would be the Gibbons-Hawking entropy\cite{i}. (3) accounts for the relation between anisotropies and the gravitational entropy. The natural choice for defining the estimator for Weyl entropy is the Weyl invariant $\mathcal{W}\equiv C_{\alpha \beta \gamma \delta }C^{\alpha \beta \gamma \delta}$, but due to isotropic singularities it is necessary to add a prefactor of the Ricci scalar $\textbf{Ric}\equiv R_{\alpha \beta }R^{\alpha \beta }$ or the Kretschmann scalar $\mathcal{K} \equiv R_{\alpha \beta \gamma \delta }R^{\alpha \beta \gamma \delta }$ . However, for the sake of convenience, we will opt for the earlier choice of using purely the Weyl invariant $\mathcal{W}$. 

\section{Naked Singularities and Weyl entropy}
The relation between the possibility of formation is proposed to be found out by finding the behaviour of the Weyl entropy in the case of a cosmological initial singularity and naked singularities from gravitational collapse. This is done by identifying the location of the singularity with respect to an outgoing geodesic -- for an initial singularity, this would be in the past, while for a gravitational collapse resulting in a singularity with an event horizon (assuming the validity of the cosmic censorship conjecture), the singularity would lie in the future for a family of geodesics that are emitted after the horizon forms -- which is the definition of a trapped surface.

The key point used in this paper is to identify the nature of the Weyl scalar based scalar $\mathcal{P}$ along an outgoing geodesic $x^{\mu }$. The Weyl curvature hypothesis states that for $x^{\mu }$ originating from an initial singularity, $\mathcal{P}(x^{\mu })$ must be diverging and monotonically increasing. In the case when we assume the cosmic censorship conjecture to be violated, a naked singularity will be found \textit{without} a trapped surface, implying that for an outgoing geodesic, the singularity lies in the past, which is similar to the cosmological initial singularity. In the first case, there is clearly a difference between the initial and the gravitational collapse singularity, whereas in the second case this is not so. In the former case, there is clearly a difference between the black hole singularity and the initial singularity, since the Weyl entropies are not the same\cite{g}. The black hole Weyl entropy would be the Hawking-Bekenstein formula, which is non-zero. We can check whether or not the two singularities are same in the latter case by looking at the gravitational entropies of the singularities. Assuming the cosmic censorship conjecture to be invalid, it would mean that a naked singularity can form and light rays will not be trapped. In this case, the similarity between the initial singularity and the naked singularity is clear, since both would be past singularities for an outgoing geodesic. As discussed previously by Barve and Singh\cite{f}, it is shown that for inhomogeneous distributions such as in Tolman-Bondi and Vaidya spacetimes, naked singularities conflict with the validity of the Weyl curvature hypothesis. This nature of diverging Weyl entropy has been studied in Barve and Singh\cite{g} is found in the case of Lemaitr\'e-Tolman-Bondi spacetimes. There are then two possibilities -- first, since the Weyl entropy must be zero \textit{at} the initial singularity\footnote{This is contradicted since the Weyl entropy diverges as you approach the past singularity in a Tolman-Bondi spacetime, which implies that the initial singularity must also have a higher Weyl entropy state than in later times.}, it is possible that naked singularities don't form, which would validate the cosmic censorship hypothesis by restricting the formation of naked singularities altogether. Second, it is possible that the Weyl curvature hypothesis may not be valid in all cases, which would have many cosmological implications. 

There is also the possibility that the very fact that there is no horizon could contribute to the invalidity of either cosmic censorship or the Weyl hypothesis, since the area of the event horizon is related to the entropy of the black hole.
\vspace{1mm}
\\The area of a black hole is related to the entropy of the black hole in the form of the Hawking-Bekenstein entropy as proposed by Hawking and Bekenstein\cite{h}:
\begin{equation}
    S_{HB}=\frac{A_{BH}}{4G\hbar }
\end{equation}
Where $A$ is the area of the event horizon and $k_{B}=c=1$. In the case when we are defining a black hole Weyl entropy, we require that $S_{W}=S_{HB}$, or more broadly $S_{W}=S_{\textrm{holographic}}$ in cases when a cosmological horizon exists, which would be the cosmological holographic principle counterpart to $S_{HB}$, called the Gibbons-Hawking entropy \cite{i}. When there does not exist a horizon, there are two implications for the Weyl entropy -- firstly, there is no horizon based contribution to $S_{W}$ as $A_{BH}=0$, and secondly, the Weyl entropy of the singularity diverges when defining the entropy density as specified by \cite{j}. That is to say that the Weyl entropy when solely considering the naked singularity may be considered to be the equivalent of the usual Weyl entropy estimator considering only the singularity and no other contribution, which is divergent. This further implies that for an outgoing geodesic in the case of a violation of the cosmic censorship conjecture, the Weyl entropy vanishes following an outgoing geodesic. This is clearly in conflict with the Weyl entropy strictly diverging along an outgoing geodesic. 

While we have not looked at other forms of gravitational entropy estimators such as the Clifton-Ellis-Tavakol proposal \cite{k} or spin coefficients proposal \cite{l}, our interest in using the purely Weyl invariant based entropy is supported by the fact that the "strong" condition here is that the Weyl invariant is a direct measure of gravitational entropy in some sense\footnote{Here, by "direct" we mean the proportionality $S_{g}\propto \mathcal{W}$. In other proposals such as the CET proposal, the measurement is done using a different approach which, while considering the Weyl tensor, are not purely independent of other factors. Interested readers are directed to the works by Chakraborty \textit{et al}\cite{c}, Pizana \textit{et al}\cite{d} and Gregoris \textit{et al}\cite{m} for cosmological studies using the CET proposal.} -- the hypothesis is simply that the Weyl entropy at the cosmological initial singularity must be zero, and it must monotonically grow, while also acting as the measure for other spacetimes where gravitational entropy is of interest such as black hole solutions. In order to capture these points, we will discuss the example of the Vaidya-de Sitter solution to the field equations, where we capture the nature of the Weyl invariant and discuss the conditions under which naked singularities pose a problem. 
\section{Charged and Uncharged Vaidya solutions}
The Vaidya\cite{n} solutions to the Einstein field equations are spacetimes characterized by imploding shells of null fluid. In the case of uncharged Vaidya spacetime to the field equations, we have a self-similar spacetime without an electric charge, while in the charged de Sitter case, we have an electric charge and a non-zero cosmological constant. In these cases, we will adopt the $(v, r, \theta , \phi )$ coordinates, where $v$ is the advanced Eddingtom time coordinate. 
\subsection{Uncharged Vaidya solution}
The Einstein field equations with a cosmological constant are given by 
\begin{equation}
    R_{\mu \nu }-\frac{1}{2}g_{\mu \nu }R+g_{\mu \nu }\Lambda =kT_{\mu \nu }
\end{equation}
In cases where we consider a de Sitter type spacetime, $\Lambda $ will not be zero. The first case is that of uncharged advanced Vaidya metric, where the electric charge and $\Lambda $ are both set to zero. Then, the Vaidya spacetime is described by the line-element:
\begin{equation}
    \mathrm{d}s^{2}=-\left(1-\frac{2m(v)}{r} \right)\mathrm{d} v^{2}+2\mathrm{d} u\mathrm{d} v+r^{2}\mathrm{d} \Omega ^{2}
\end{equation}
For this to be self-similar, we impose the condition of $m(v)=\chi _{0}v$, so that $\chi _{0}$ is the infall rate of null dust\cite{o}. For any value less than or equal to $0.125$, a naked singularity is formed. The Weyl invariant $\mathcal{W}$ is a function of $v$ and $r$, given by
\[\mathcal{W}=\frac{12\chi ^{2}v^{2}}{r^{6}}\]
We define the tangent of an outgoing geodesic from the naked singularity as $X=v/r$, with $X_{0}$ being the limiting value. Further, the Ricci scalar is zero in this case, indicating a dominance of the Weyl tensor. We see that the Weyl invariant diverges as $r\to 0$, which is a conflicting statement with the Weyl curvature hypothesis, which states that the initial singularity must have a lower entropy value that diverges with time\cite{f}. Therefore, the similarity between the naked singularity here and the initial singularity is in conflict with the Weyl curvature hypothesis. 
\subsection{Charged Vaidya-de Sitter solution}\label{sec:cvds}
Another case of Vaidya spacetimes is that of those with charge and $\Lambda $ being non-zero. Such a spacetime is a Vaidya-de Sitter spacetime with charge, with the metric
\begin{equation}\label{eq:cvds}
    \mathrm{d}s^{2}=-\left(1-\frac{2m(v)}{r}+\frac{q^{2}(v)}{r^{2}}-\frac{\Lambda r^{2}}{3} \right)\mathrm{d}v^{2}+2\mathrm{d}u\mathrm{d}v+r^{2}\mathrm{d}\Omega ^{2}
\end{equation}
This spacetime represents inflow of null fluid with a non-zero charge in an empty de Sitter region. Assuming the collapse takes places at $r=0$ with the last infalling cloud at some time $v=t$, we find that there are three conditions on $m(v)$ and $q(v)$ being proportional to $v$. The bounds for this function are $v<0$, $v\in [0, t]$ and $v>t$ respectively \cite{p}. Based on this, we have conditions on $m(v)\propto v$ and $q^{2}(v)\propto v^{2}$, using which the piecewise definitions of $m$ and $q$ can be found out. In this case, for the interval $v\in [0, t]$, we define $m(v)=\lambda v$ for $\lambda >0$, while after $v=t$, $m(v)$ is a constant $M$. A similar definition is used for $q(v)$, where we define $q(v)=\theta v$ for $v\in [0, t]$ with $\theta >0$ and a constant $\Theta $ for $v>t$. For $v<0$, both $m(v)$ and $q(v)$ are zero. The factors $m(v)$ and $q(v)$ are determined in accordance to the energy conditions, and for the bound $[0, t]$, this would be the charged Vaidya-de Sitter spacetime. In order to see the nature of the Weyl invariant for this spacetime, we start by defining the tangent to an outgoing geodesic by $X=v/r$. Using these definitions, we can show that for null geodesics, 
\begin{equation}\label{eq:cvdssc}
    \frac{\mathrm{d}r}{\mathrm{d}v}=\frac{1}{2}\left(1-2\lambda X+\theta ^{2}X^{2}-\frac{\Lambda r^{2}}{3} \right)
\end{equation}
This clearly has a singularity at $r, v=0$. In order to determine the nature of the singularity, we would need to check if there are non-spacelike curves that are escaping from the singularity. In order to do so, we let a limiting value $X_{0}$ on $X$:
\begin{equation}\label{eq:cvdslv}
    X_{0}=\lim _{r, v\to 0}\frac{v}{r}
\end{equation}
By differentiating the numerator and denominator and using \eqref{eq:cvdssc}, \eqref{eq:cvdslv} can be solved to yield
\begin{equation}\label{eq:cvdsr}
    \theta ^{2}X_{0}^{3}-2\lambda X_{0}^{2}+X_{0}-2=0
\end{equation}
If this has a positive root, it implies that there exists a locally naked singularity, which \eqref{eq:cvdsr} does due to the non-negative values of $\lambda $ and $\theta $. This can be seen by observing that \eqref{eq:cvdsr} is a cubic equation, and therefore it must be true that $\theta  >0$ and $\lambda >0$. 

The Weyl invariant $\mathcal{W}$ is a function of $X$ and $r$, which can be found out to be
\begin{equation}\label{eq:wicvds}
    \mathcal{W}(X, r)=\frac{48}{r^{4}}\left(\lambda ^{2}X^{2}-2\lambda \theta ^{2}X^{3}+\theta ^{4}X^{4} \right)
\end{equation}
Unlike the uncharged Vaidya spacetime case, this spacetime has a non-dominant Weyl scalar over $\mathbf{Ric}$, which diverges at the same rate as the Weyl invariant. In this case, equation \eqref{eq:wicvds} shows that the Weyl invariant diverges at the singularity. Since this solution allows naked singularities and has a diverging Weyl invariant at the singularity, it may be considered that the Weyl entropy for an outgoing geodesic from the singularity has a lower entropy state further in time than at the singularity. In contrast, an outgoing geodesic from the cosmological initial singularity would yield a diverging Weyl entropy. Therefore, this may be considered as a contradictory statement to the Weyl curvature hypothesis assuming cosmic censorship is invalid.
\subsection{Charged Vaidya anti-de Sitter solution}
If $\Lambda <0$, the metric \eqref{eq:cvds} represents a charged Vaidya solution in an AdS background\cite{lemos1999}. In this case too, a similar approach can be used to find out the nature of the singularity for specific initial data\cite{0106060} and the Weyl invariant. By following the same approach as the previous case as illustrated in section \hyperref[sec:cvds]{3.2}, we start by defining the limiting value $X_{0}$ as $X_{0}=\lim _{r, v\to 0}\; \frac{v}{r}$. 
By differentiating the numerator and the denominator, we can use the slope condition for this spacetime to get \eqref{eq:cvdsr}, which in this case would have varying values of $\theta $ and $\lambda $ with respect to $\Lambda $ that admit a positive root, which implies the existence of a naked singularity. 
\section{Tolman-Bondi solutions}
As further examples on the diverging Weyl entropy, consider the case of the spherical dust collapse from the Tolman-Bondi solutions. We will discuss the Tolman-Bondi dust collapse under asymptotically flat ($\Lambda =0$) and asymptoticaly de Sitter ($\Lambda >0$) spacetimes. The general form of the metric for such cases is given by
\begin{equation}\label{eq:gtb}
    \mathrm{d}s^{2}-\mathrm{d}t^{2}+e^{-2\chi (r)}\mathrm{d}r^{2}+R(r, t)\mathrm{d}\Omega ^{2}\;\;\;.
\end{equation}
\subsection{$\Lambda =0$}
Another spacetime with a diverging Weyl entropy in a gravitational collapse is that of Tolman-Bondi spacetime, defined by the metric
\begin{equation}
    \mathrm{d}s^{2}=-\mathrm{d}t^{2}+\frac{R'(r, t)^{2}}{1+f(r)}\mathrm{d}r^{2}+R(r, t)^{2}\mathrm{d}\Omega ^{2}
\end{equation}
where the prime denotes differentiation w.r.t. the comoving $r$ coordinate. Defining the mass function for the cloud of dust as $m(r)$, the Weyl invariant is of the form
\begin{equation}
    \mathcal{W}(r, t)=\frac{48}{R(r, t)^{4}}\left(\frac{m(r)}{R(r, t)}-\frac{m'(r)}{3R'(r, t)} \right)^{2}
\end{equation}
This clearly diverges at $R=0$ -- for both ingoing and outgoing geodesics, there is a divergent Weyl invariant. In this case, $\mathbf{Ric}$ diverges at the same rate as that of $\mathcal{W}$. In order to find the condition for naked singularities as seen before, we define the tangent $R=Xr^{\alpha }$ for $\alpha >1$. For this spacetime too, we see that the Weyl invariant is a function of $r$ such that at $r=0$ there is a naked singularity for the range $\alpha \in [5/3, 3]$. Clearly again, along an outgoing geodesic we observe a vanishing Weyl invariant, which implies that the naked singularity cannot be similar to the initial singularity, in conflict with the Weyl curvature hypothesis\cite{g}. 
\subsection{$\Lambda >0$}
This is the case of Tolman-Bondi dust scenario \eqref{eq:gtb} seen above in asymptotically de Sitter spacetime. The stress energy momentum tensor is given by
\begin{equation}
    T_{\mu \nu }=\rho (r, t)u_{\mu }u_{\nu }-g_{\mu \nu }\frac{\Lambda }{8\pi }
\end{equation}
And we have $m(r, t)=\frac{R}{2}\left(\Dot{R}^{2} -\frac{\Lambda R^{2}}{3}+f\right)$ In this case the constraint on $R$ is that it must satisfy
\[m'=4\pi \rho R^{2}R'\]
The Kretschmann scalar $\mathcal{K}$ diverges, as does the Ricci scalar \textbf{Ric}. The determination of a singularity's nature being naked in this case too can be seen by finding an outgoing null geodesic that has an origin at the singularity. At $r=0$, we define $\rho (r, t=0)$, and by defining the expansion as proposed by Barve and Singh\cite{f}, we find that there exist initial data\footnote{The interested reader is directed to Goncalves\cite{r} for a complete derivation on the initial data and the \textit{locality} of the singularity, where the expansion put forward by Barve and Singh\cite{g} is used.} for which there exists \textit{at least} one outgoing null geodesic with a starting point at the singularity. Further, the Weyl invariant is found to be similar to that of the $\Lambda =0$ case, and at the singularity for specific initial data, $\mathcal{W}$ diverges to infinity. We have discussed subcases of two general solutions to the field equations, the Vaidya and the Tolman-Bondi collapse. In the next section, we will briefly remark on the nature of gravitational entropy as a horizon based entropy. 
\section{Horizon based entropy}
By definition, the gravitational entropy of a spacetime with a horizon must reduce to its corresponding horizon based entropy, i.e. 
\[S_{\text{grav}}\thicksim A_{\text{horizon}}\]
Then, we can base the definition of gravitational entropy on the horizon entropy. This was proposed by Rudjord and Gron \cite{j} and verified in several literatures (such as the papers on black holes\cite{q}, wormholes\cite{e}, etc.) for various solutions. However, in the case of Petrov type \textbf{O} spacetimes with a cosmological horizon, it fails to recover the appropriate entropy, which is the Gibbons-Hawking entropy. Since $S_{\text{grav}}\thicksim C_{\alpha \beta \gamma \delta }C^{\alpha \beta \gamma \delta }$, the gravitational entropy must be zero, indicating a purely thermodynamic entropy contribution. This is not a serious problem -- for instance, by use of spin coefficients as proposed by Gregoris and Ong\cite{l}, it is possible to show that a horizon based entropy can be defined for cases where $C_{\alpha \beta \gamma \delta }=0$. Therefore, while the Weyl estimator works in some cases, it is possible that there are other estimators that define gravitational entropy more "strongly"\footnote{In the spin coefficients approach, this refers to the use of the Weyl tensor implictly, since the ground for defining gravitational entropy is no longer a direct quantification of the Weyl tensor but instead the use of spin coefficients takes its place, which reduces to an appropriate form of the Weyl tensor when it is non-zero; refer to sections II-IV in the paper by Gregoris and Ong\cite{l}.} -- in any case, since the horizon based entropy is defined on the pretext of the spacetime having a horizon, it is clear that if a gravitational collapse leads to a naked singularity, the Weyl entropy there would face a divergence due to the singularity in the volume integral. This can be shown by considering the direct Weyl approach\cite{j}, where we define the volume integral as
\begin{equation}
    S_{grav}=k_{s}\int _{V}\; \frac{C_{\alpha \beta \gamma \delta }C^{\alpha \beta \gamma \delta }}{\mathcal{R}} dV
\end{equation}
Where $k_{s}$ is a constant and $\mathcal{R}$ is a curvature invariant correction to the Weyl invariant due to rescaling for isotropic singularities\cite{Wainwright}, which further also indicates the dominance of the Weyl invariant over the Riemann tensor based invariants $\mathcal{R}=\mathcal{K}\equiv R_{\alpha \beta \gamma \delta}R^{\alpha \beta \gamma \delta }$ or $\mathcal{R}=\mathbf{Ric}\equiv R_{\alpha \beta }R^{\alpha \beta }$. Clearly, in order to define the horizon entropy of a black hole, we must subtract the singularity by defining a spherical element of radius $\epsilon \to 0$ and subtracting this from the singularity. However, in the case of naked singularities, no horizon is observed -- then, a divergent Weyl invariant directly implies a diverging gravitational entropy, which is the base of the conflict between cosmic censorship and the Weyl curvature hypothesis. Since the constraint to begin with on gravitational entropy is that it must be independent of local matter fields $\mathbf{\Psi }$ on $(M, g)$, the Weyl invariant is the appropriate choice. Further, the validity of cosmic censorship, even if found to be true, does not guarantee the validity of the Weyl curvature hypothesis in the Weyl entropy-cosmic censorship scheme. Since the diverging nature of the Weyl invariant is geometrically intrinsic to singularities, the validity of the Weyl curvature hypothesis with respect to cosmic censorship remains to be answered. 

We can note that the validity of the Weyl curvature hypothesis has an interesting place in cosmology. In terms of cosmology, it would be preferable that the Weyl curvature hypothesis is true -- however, certain cases point towards the contrary, such as Gregoris \textit{et al}\cite{m}. In addition to this, the vanishing Weyl tensor for the de Sitter spacetime forces us to question the Weyl curvature hypothesis. The cosmological no hair conjecture states that certain cosmologies are such that in future states the expansion rate $a(t)$ becomes asymptotically de Sitter, i.e. 
\begin{equation}
    \lim _{t\to \infty } a(t)_{M}=\exp{Ht}
\end{equation}
These certain cosmologies include the Bianchi spacetimes (excluding the $IX$ case)\cite{z}, under the requirement of the satisfaction of the strong energy condition
\[\left(T_{\mu \nu }-\frac{1}{2}g_{\mu \nu }tr(T_{\mu \nu })\right)\geq 0\]
The cosmological no hair conjecture can be stated as: "\textit{for a cosmology $(M, g)$ with a positive cosmological constant $\Lambda $ and spatial curvature $k\leq 0$, the future state would be asymptotically de Sitter}". Due to this, the future state of the cosmology would have a Weyl curvature that is asymptotically zero, conflicting with the Weyl curvature hypothesis. Due to this, a firm choice of the validity of the Weyl curvature hypothesis cannot be provided -- the possible relation to cosmic censorship described previously is based on the assumption that the Weyl curvature hypothesis is true and that along all outgoing geodesics the Weyl curvature is strictly increasing -- if the Weyl curvature hypothesis is proved to be true, it would mean that cosmic censorship would hold via the previously discussed relation.
\section{Remarks}
The validity of the Weyl curvature hypothesis would have implications on the cosmic censorship conjecture in its weak form. The diverging nature of the Weyl invariant poses a problem with the naked singularity after the collapse -- this results in two possible propositions: (1) that either the Weyl invariant cannot be a suitable estimator of gravitational entropy directly, or (2) the Weyl entropy in itself is not suitable for being related to the singularity. However, the Weyl invariant has been found to be conveniently straightforward for determining gravitational entropy -- for example, for black hole solutions it results in the correct entropy measure, i.e. the Hawking-Bekenstein entropy. However, this comes at the price of being fully diverging in the case of singularities without a horizon for entropy measure, which we would ordinarily expect for curvature singularities. The second possibility, that the Weyl entropy is not suitable in itself has not been considered because the natural choice of validity in view of cosmology is that the Weyl curvature hypothesis is true, i.e. the Weyl entropy along outgoing geodesics from the initial singularity monotonically increases. Therefore, if cosmic censorship is found invalid, it might imply that the Weyl entropy vanishing along outgoing geodesics with past endpoint at the singularity formed by gravitational collapse is naturally allowed. On the other hand, it could be possible that Weyl entropy is allowed to be reducing along outgoing geodesics originating from \textit{local} singularities, i.e. singularities from which only one set of null outgoing geodesics escape. However, in many cases this is not so -- global singularities also occur in the spacetime, which complicates the situation further. It must be remarked that while the discussion in this paper does not take further technicalities of the weak cosmic censorship, it must be noted that under certain conditions global naked singularities do not occur -- refer to Christodoulou's work\cite{s, t, u} for a view on the nature of matter content and the "generic" nature of naked singularities to take into account of conditions on the physical nature of matter in the spacetime such as the satisfaction of the dominant energy condition, etc. It must be mentioned that self-similar spacetimes pose a problem with determining the gravitational entropy along homotheticities the form $\mathcal{W}/\mathbf{Ric}$ as it was shown by Pelavas and Lake\cite{PL2018}. However, for the consideration of gravitational entropy in this article, we have solely looked at the involvement of $\mathcal{K}$ and $\mathbf{Ric}$ without the involvement of $\mathcal{R}$ in the gravitational entropy estimator. Further, whether there exists a possible relation between the strong cosmic censorship and the Weyl curvature hypothesis is yet to be studied.

\end{document}